# Long-Baseline Neutrino Oscillation Experiments

*Alexandre B. Sousa*

*Department of Physics, Harvard University, Cambridge, Massachusetts 02138, USA, asousa@physics.harvard.edu*

**Abstract**

During the past decade, long-baseline neutrino experiments played a fundamental role in confirming neutrino flavor change and in measuring the neutrino mixing matrix with high precision. This role will be amplified with the next generation of experiments, which will begin probing the possibility of *CP* violation in the leptonic sector and possibly pin down the neutrino mass hierarchy. An account of the most recent results from the MINOS experiment is presented, along with the earlier measurement from the K2K experiment. The next generation projects, T2K and NOνA, are described and their current status, schedule, and physics reach discussed. Finally, we report on future efforts, currently in the R&D stage, such as the LBNE and T2KK projects.

## 1. Introduction

There is now compelling evidence that electron and muon neutrinos change flavor as they propagate from their production point. This phenomenon was first established by observation of the disappearance of muon neutrinos originating in the atmosphere [1, 2], and of electron neutrino fluxes from the Sun [3, 4] and from terrestrial reactors [5]. However, the full determinations and precise measurement of neutrino oscillation parameters is only possible using a man-made high intensity neutrino source, such as an accelerator-generated neutrino beam. The beam is characterized near the production target using beam monitors and possibly a near detector, and then measured again at a far location, distant enough for the neutrinos to have undergone flavor change. This is the basic idea behind most of the current and future neutrino oscillation experiments, to be detailed below.

## 2. Neutrino Oscillations

The idea that neutrinos may oscillate, originally introduced by Pontecorvo [6] is based on the hypothesis that the three known neutrino weak (flavor) eigenstates $\nu_e$, $\nu_\mu$, $\nu_\tau$, do not coincide with the neutrino mass eigenstates $\nu_1$, $\nu_2$, $\nu_3$. In fact, the flavor states $\nu_\alpha$ can be represented as linear combinations of the mass states, $|\nu_\alpha\rangle = \sum_{i=1}^{3} U^*_{\alpha i}|\nu_i\rangle \quad (\alpha = e,\mu,\tau)$, where $U$ is the PMNS unitary matrix [7]. In a simplified scenario including only two weak eigenstates related to two mass eigenstates, the survival probability of a neutrino of energy $E_\nu$ and flavor $\alpha$ after propagating a distance $L$ in vacuum is:

$$P(\nu_\alpha \to \nu_\alpha) \simeq 1 - \sin^2 2\theta_{ij} \sin^2\left(1.27\Delta m^2_{ij}(\mathrm{eV}^2)\frac{L(\mathrm{km})}{E_\nu(\mathrm{GeV})}\right),$$

where $\Delta m^2_{ij}$ is the difference of the squares of the neutrino masses $m_j^2 - m_i^2$, and $\theta_{ij}$ is the neutrino mixing angle. For the three known neutrinos, the mixing matrix is expressed in terms of the rotation angles $\theta_{12}$, $\theta_{23}$, $\theta_{13}$, and a *CP*-violating phase $\delta$ by:

$$\underset{(a)}{\begin{bmatrix} 1 & & \\ & c_{23} & s_{23} \\ & -s_{23} & c_{23} \end{bmatrix}} \underset{(b)}{\begin{bmatrix} c_{13} & & s_{13}e^{-i\delta} \\ & 1 & \\ -s_{13}e^{i\delta} & & c_{13} \end{bmatrix}} \underset{(c)}{\begin{bmatrix} c_{12} & s_{12} & \\ -s_{12} & c_{12} & \\ & & 1 \end{bmatrix}}$$

Here, $c_{ij} \equiv \cos\theta_{ij}$ and $s_{ij} \equiv \sin\theta_{ij}$. A nonzero value of $\delta$ allows the possibility of *CP* violation in the leptonic sector, provided that $\sin\theta_{13}$ is also nonzero. In this factorized representation of the mixing matrix, the atmospheric term (*a*) is determined by atmospheric and long-baseline accelerator neutrino experiments, the solar term (*c*) is determined by solar and long-baseline reactor experiments, and the mixed term (*b*) is determined by long-baseline accelerator and short-baseline reactor experiments. The description of the phenomenology of oscillations among three active neutrinos is completed by the two independent mass-squared differences $\Delta m^2_{21}$ and $\Delta m^2_{32}$. The solar neutrino data from SNO [4] and reactor neutrino data from KamLAND [5] determine the solar mixing angle to be $\theta_{12} = 34.1^{+1.16}_{-0.84}$ deg, and the solar mass-square difference to be $\Delta m^2_{21} = (7.59 \pm 0.21) \times 10^{-5} \mathrm{eV}^2$. From the atmospheric neutrino data [2], a lower limit of $\sin^2 2\theta_{23} > 0.96$ (90% C.L.) is obtained, allowing the possibility of maximal mixing, and a measurement of $|\Delta m^2_{32}| = 2.19^{+0.14}_{-0.13} \times 10^{-3} \mathrm{eV}^2$ is made. The reactor data from CHOOZ [8] provides the limit $\sin^2 2\theta_{13} < 0.15$ (90% C.L.). There is no present experimental constraint on the value of $\delta$. Furthermore, there is no indication on the sign of $\Delta m^2_{32}$, also referred to as neutrino mass hierarchy, which defines whether $m_3 > m_2$ or $m_3 < m_2$.

## 3. Long-Baseline Neutrino Experiments

Conceptually, a long-baseline neutrino oscillation experiment employs an intense $\nu_\mu$ beam (> 100 kW beam power) measured at a near site $\mathcal{O}$(1 km) downstream of beam production, and measured again at a far site placed at a distance $\mathcal{O}$(100 km) away, to study neutrino oscillations. Neutrino flavor change is studied in long-baseline experiments either by measuring the disappearance of the muon neutrino flux or by looking for appearance of $\nu_e$ or $\nu_\tau$ events.

The first long-baseline experiment was K2K [9], followed by MINOS [10] and OPERA [11], and the next-generation experiments T2K [12] and NOvA [13]. K2K was designed to verify the atmospheric neutrino oscillation results of the SuperKamiokande experiment; MINOS is aimed at precisely measuring the atmospheric oscillation parameters; OPERA hopes to obtain direct confirmation of $\nu_\mu \rightarrow \nu_\tau$ oscillations; T2K is designed to observe $\nu_\mu \rightarrow \nu_e$ and make the first measurement of $\theta_{13}$; Besides measuring $\theta_{13}$, NOvA may also be able to determine the neutrino mass hierarchy, due to a longer baseline than T2K's. A summary of the characteristic parameters of long-baseline experiments is presented in Table I.

| Exp. | Run | Peak $E_\nu$ (GeV) | Baseline (km) | Det. Tech. | FD Mass (kton) |
|---|---|---|---|---|---|
| **K2K** | 1999-2004 | 1.0 | 250 | $H_2O$ Č | 50 |
| **MINOS** | 2005-2011? | 3.3 (LE) | 735 | Steel+ Scint. | 5.4 |
| **OPERA** | 2008- | 17 | 732 | Emulsion | 1.25 |
| **T2K** | 2010- | 0.6 | 295 | $H_2O$ Č | 50 |
| **NO$\nu$A** | 2013- | 2.0 | 810 | Liq. Scint. | 14 |
| **LBNE** | 2020? | 1-10 | 1300 | $H_2O$ Č Liq. Ar | 300/50 |

***Table I:*** *The main parameters characterizing past, current and future long-baseline experiments.*

### 3.1. Building a Neutrino Beam

The neutrino beams used in long-baseline neutrino experiments are all created in very similar ways. A beam of protons as intense as possible is focused on a target, usually made of graphite, producing secondary pions and kaons. The emerging mesons are focused by parabolic magnetic horns and allowed to decay into muons and muon neutrinos over several hundred meters. The muons are absorbed in the surrounding rock and used to monitor the beam intensity and position, whereas the tertiary neutrinos continue through to the near and far sites of the experiments. Typically, the magnetic horns are optimized to focus the secondary $\pi^+$. A muon antineutrino beam may be created by inverting the current in the horns and preferentially focusing $\pi^-$. The $\nu_e / \bar{\nu}_e$ contamination in the beam, resulting from $K^0$ and $\mu^+$ decays, is typically ~1%. A depiction of the different components of a neutrino beam line is shown in Figure 1.

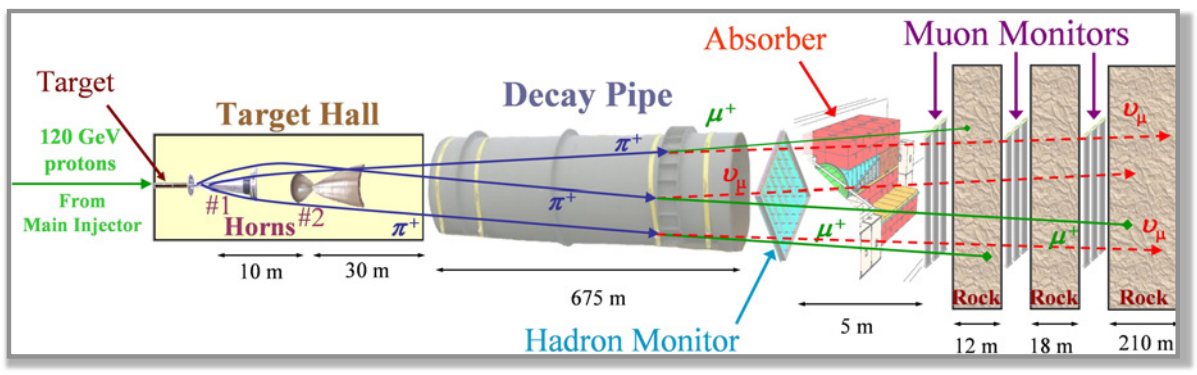

*Fig. 1: Schematic diagram displaying the components involved in creating the NuMI neutrino beam.*

## 3.2. The K2K Experiment

The KEK to Kamioka experiment was the first long-baseline neutrino experiment. It ran from 1999 until 2004 with the objective of independently confirming the observation of neutrino oscillations by the Super-Kamiokande atmospheric neutrino experiment. The muon neutrino beam was created from 12 GeV protons accelerated by the KEK proton synchrotron facility (KEK-PS) with an average intensity of $5 \times 10^{20}$ protons per pulse and an average neutrino energy of 1.3 GeV. K2K measured the beam with two near detectors, a 1 kton water Čerenkov tank and a lead glass calorimeter updated to a fine grained scintillator tracker (SciBar), and with one far detector, the 50 kton (22.5 kton fiducial) water Čerenkov tank used by Super-Kamiokande, placed 250 km away. In the full data sample collected, corresponding to $1.0 \times 10^{20}$ protons on target (POT), K2K observed 112 neutrino candidate events at the far detector, against an expectation of $158.1^{+9.2}_{-8.6}$ for null oscillations, which are thus excluded at $4.3\sigma$ [11]. Of the total number of events observed, 58 display a single muon-like Čerenkov ring, resulting from quasi-elastic scattering interactions $\nu_\mu + n \rightarrow \mu^- + p$. The reconstructed neutrino energy spectrum of the quasi-elastic events selected in the far detector is shown in Figure 2. The reconstructed energy spectrum displays an energy-dependent distortion, characteristic of neutrino oscillations, consistent with the oscillation parameter values of $\Delta m^2 = 2.8^{+0.7}_{-0.9} \times 10^{-3}$ eV$^2$ and $\sin^2 2\theta = 1.0$. The 90% C.L. allowed regions in oscillation parameter space are shown in Figure 3.

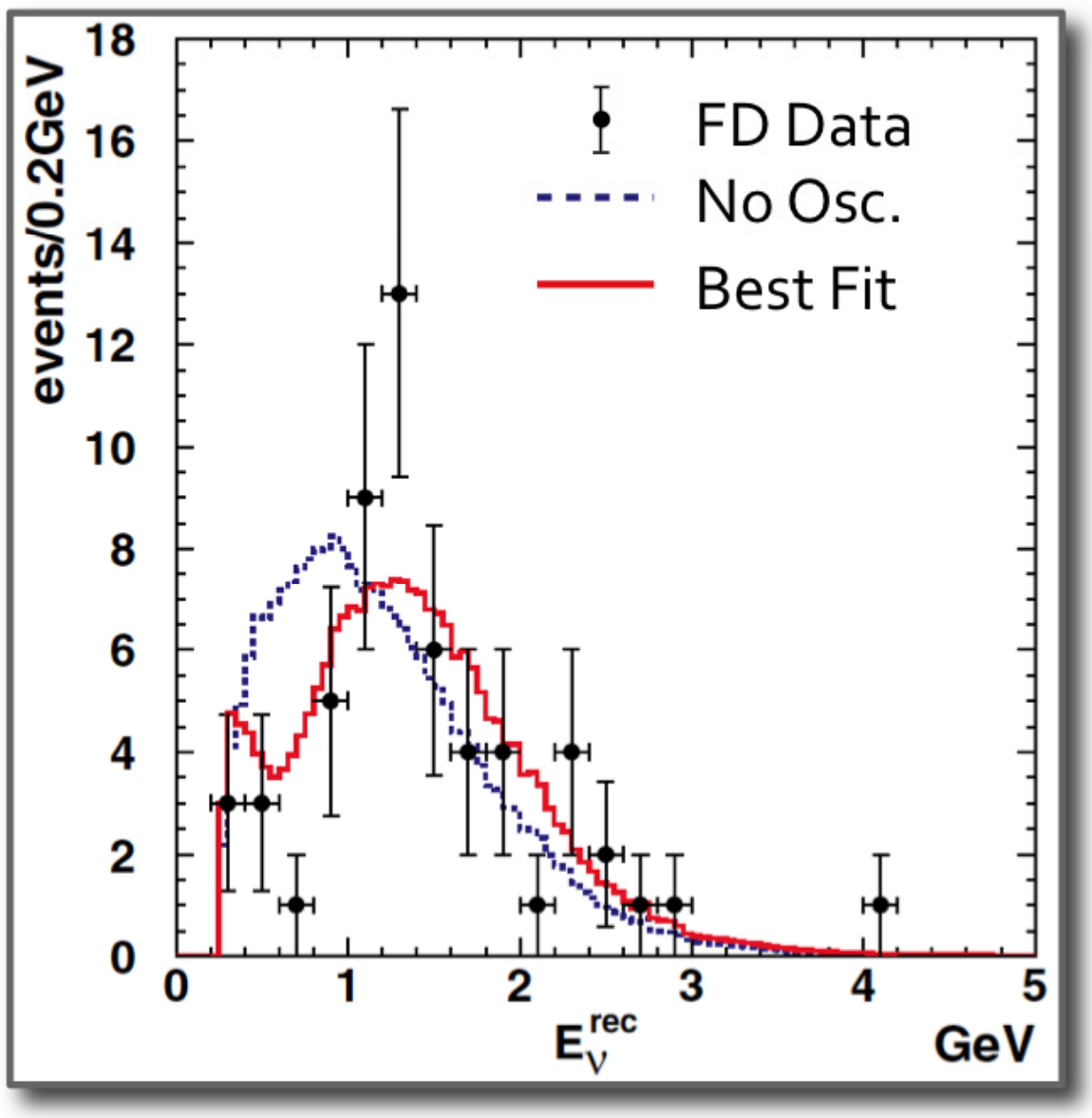

*Fig. 2: Reconstructed neutrino energy spectrum of the 58 quasi-elastic neutrino interaction candidates observed by K2K, shown along with the null-oscillation expectation (dashed blue line) and the best oscillation fit to the data (solid red line). The blue and red histograms are normalized to the data.*

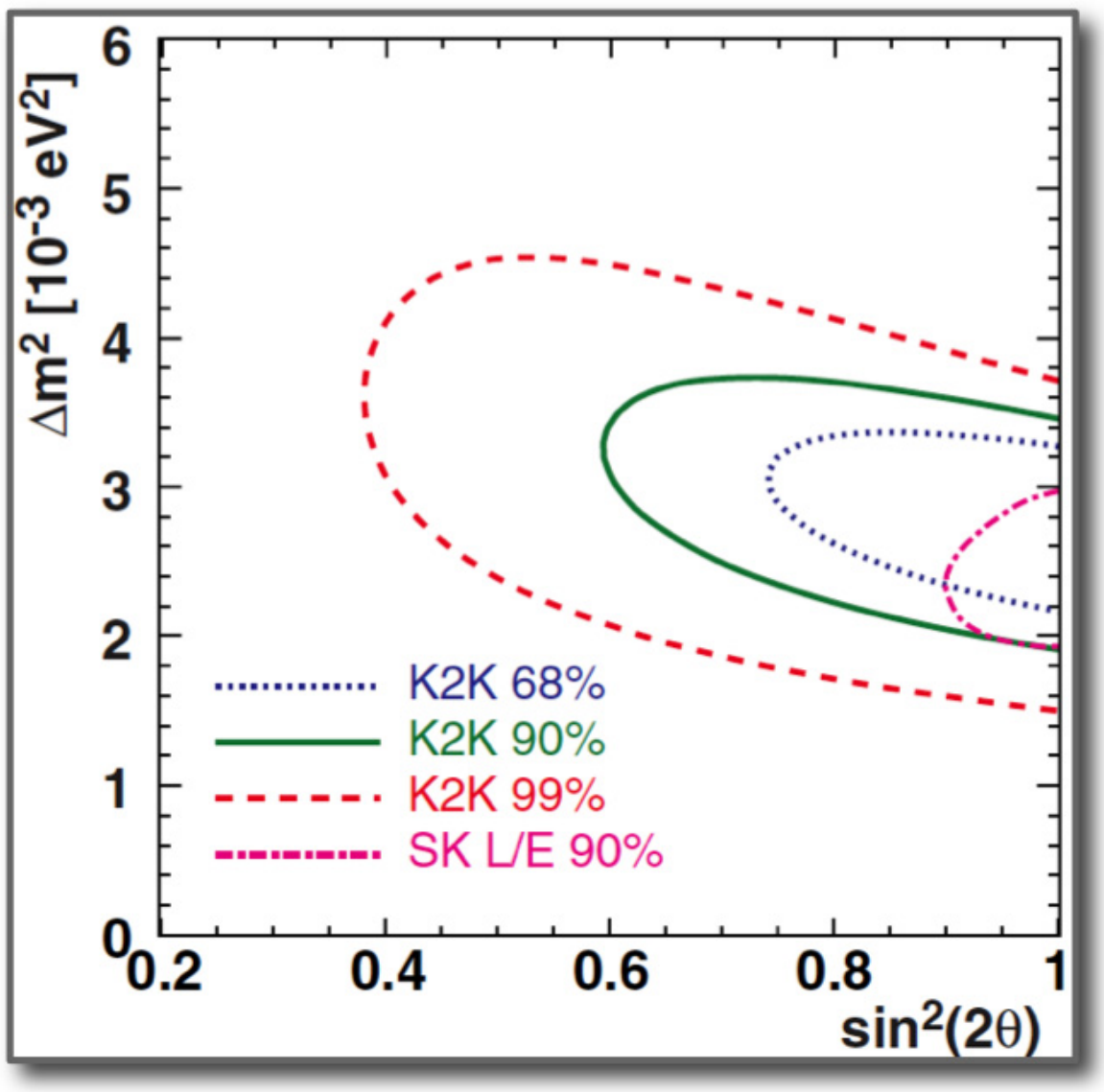

*Fig. 3: Comparison of 68%, 90% and 99% C.L. allowed regions from oscillation fits to K2K data with the 90% C.L. contour from Super-Kamiokande.*

## 3.3. The MINOS Experiment

The Main Injector Neutrino Oscillation Search experiment has as primary goal to make the first precision measurements of the atmospheric oscillation parameters by greatly improving on the statistics of the collected data sample. It uses the Neutrinos from the Main Injector (NuMI) neutrino beam created from 120 GeV protons from the Main Injector facility at Fermilab. The beam operates with 300 kW average beam power and has a peak energy of ~3 GeV in the low energy (LE) configuration. The current energizing the two magnetic horns may be inverted to focus $\pi^-$ and create an antineutrino beam. The beam is sampled at two locations; a 1 kton near detector at Fermilab, placed 1 km downstream of the target; and a 5.4 kton far detector, located 735 km away in the Soudan mine, in northern Minnesota. The detectors are magnetized planar steel/scintillator calorimeters and are functionally identical, allowing for extensive cancellation of systematic uncertainties associated with detector response, interaction cross sections and neutrino beam flux. The MINOS results presented here are obtained from data recorded between 2005 and 2010, corresponding to an exposure of $7.2\times10^{20}$ POT in neutrino running mode and $1.7\times10^{20}$ POT in antineutrino mode. The events observed in the detectors have distinct topologies depending on the interaction type: $\nu_\mu$ charged current (CC) events display a long muon track along with hadronic activity near the event vertex; neutral current (NC) events tend to be short and contain a diffuse hadronic shower; and $\nu_e$ CC events are also short and contain a compact shower displaying a typical electromagnetic profile. The MINOS CC disappearance analysis selects $\nu_\mu$ CC using a k-nearest neighbor technique using variables related to the muon track properties [12]. The neutrino energy is reconstructed by summing the muon track momentum (determined from range or curvature measurements) and the hadronic shower energy. A total number of 1986 events is observed in the far detector, whereas the expectation without oscillations is 2467 events. The energy spectrum of selected events is shown in Figure 4. The data is best fit by the neutrino oscillation hypothesis. The oscillation fit uses the two-neutrino flavor approximation for the survival probability shown earlier. The values of the oscillation parameters obtained from the fit are $|\Delta m^2_{32}|=2.35^{+0.11}_{-0.08}(\text{stat.+syst.})\times10^{-3}\ \text{eV}^2$ and $\sin^2 2\theta_{23}>0.91$ (90% C.L.) [14].

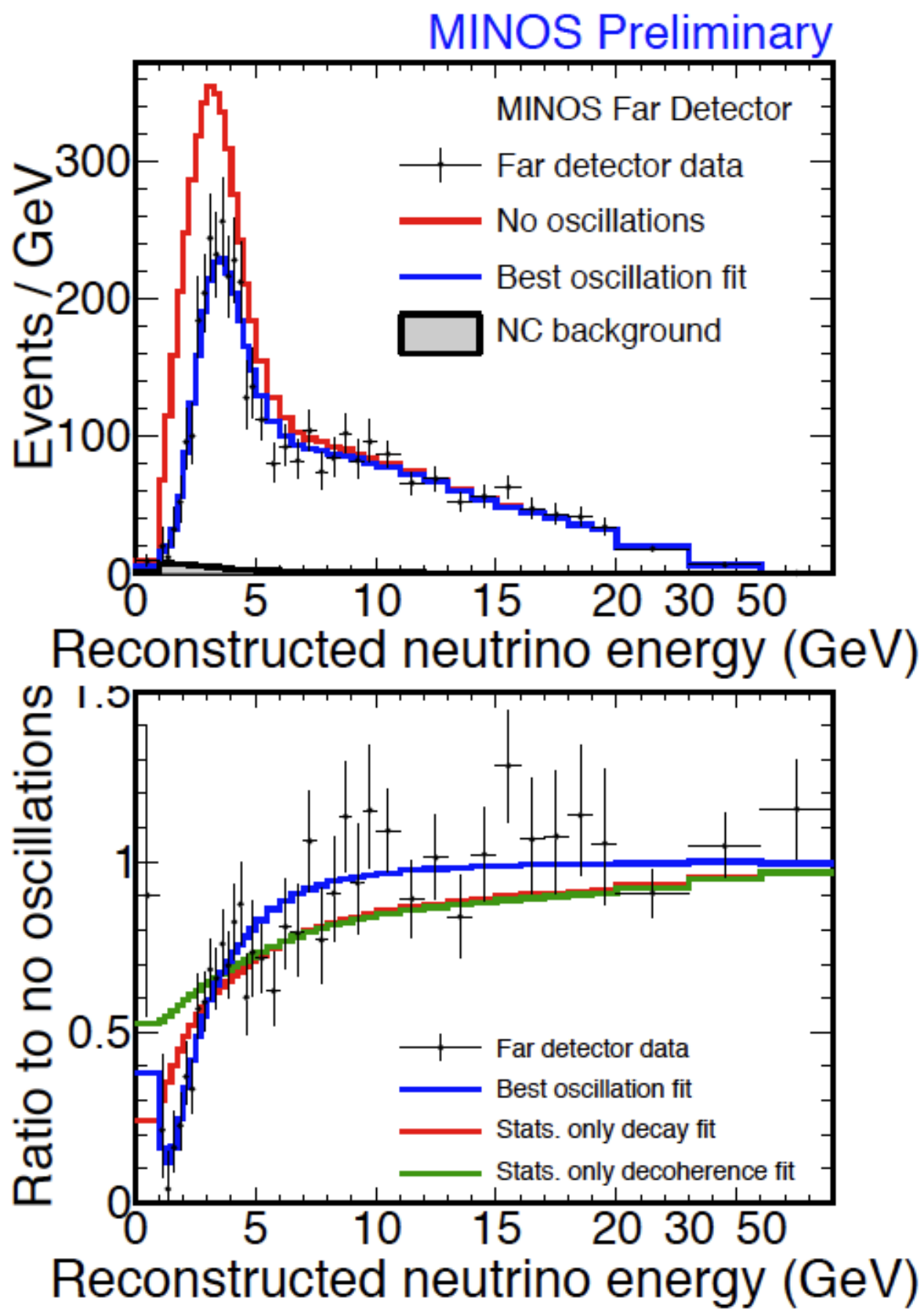

*Fig. 4: (Top) Reconstructed energy spectrum of $\nu_\mu$ CC events selected in the MINOS far detector. Black points are the data, the red histogram shows the prediction from near detector measurements, assuming no oscillations, and the blue histogram shows the best oscillation fit to the data. (Bottom) Ratio of far detector energy spectrum to the null oscillation prediction. Black points show the data, the blue line represents the best oscillation fit, the red line depicts the fit from a pure decay model, and the green line shows the fit from a pure decoherence model.*

Figure 4 also displays the ratio of the observed energy spectrum to the prediction for null oscillations, the best oscillation fit and fits to alternative models, pure decoherence [15] and pure decay [16]. Pure decoherence is disfavored relative to oscillations at more than 8σ, whereas pure decay is disfavored at more than 6σ. The allowed regions around the best fit values of the oscillation parameters are shown in Figure 5.

MINOS is able to probe the possibility of oscillations into sterile neutrinos by measuring NC events in the far detector [17]. The NC interaction rate is not affected by oscillations among the three known neutrino flavors. However, if oscillations into one or more sterile species were to occur, an energy-dependent depletion of the NC event rate would be observed.

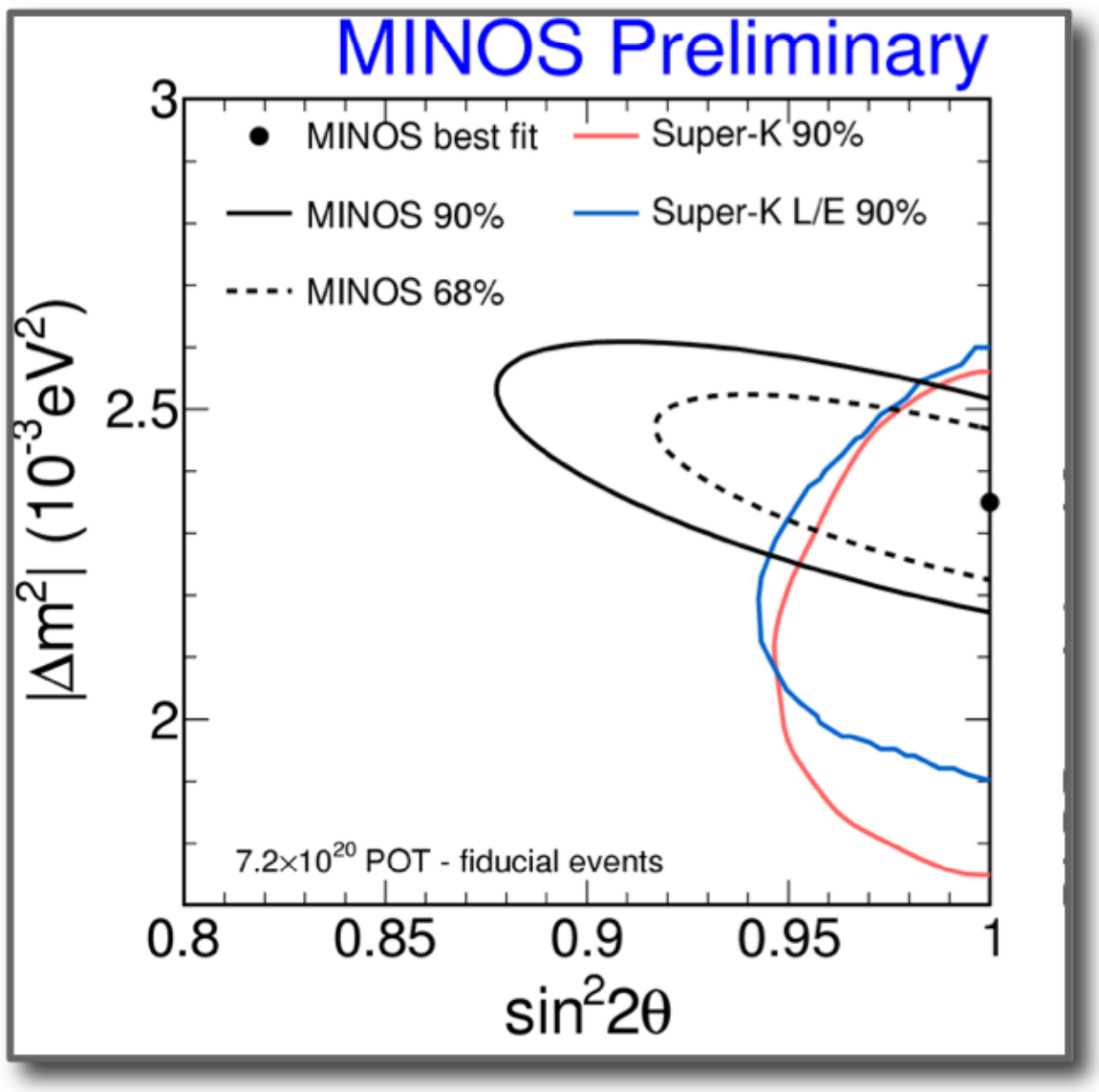

***Fig. 5:*** *The MINOS 68% C.L. and 90% C.L. allowed regions for the oscillation parameters. Also shown are the contours from the zenith angle (red line) and L/E (blue line) Super-Kamiokande analyses.*

The MINOS NC analysis employs cuts on event length and the length of the reconstructed track relative to the length of the reconstructed shower to select shower-like events. The main background affecting this selection arises from highly inelastic CC interactions for which the muon track is buried in the hadronic shower. Furthermore, the analysis accepts any $\nu_e$ CC events appearing in the far detector, requiring the interpretation of results to take into account the possibility of $\nu_e$ appearance. The analysis selected 802 NC event candidates, whereas 757 events are predicted in the absence of oscillations. No depletion of the NC event rate is observed. The reconstructed energy spectrum of the selected events is displayed in Figure 6. From fits to the data using scenarios that include sterile neutrino mixing, the fraction of active neutrinos that may oscillate into sterile neutrinos is constrained to be less than 22% assuming no $\nu_e$ appearance, and less than 40% (with $\nu_e$ appearance), both at 90% C.L. [14].

MINOS has also attempted to measure the $\theta_{13}$ mixing angle by searching for $\nu_e$ appearance in the far detector. The $\nu_e$ candidate events are selected by identifying an electromagnetic shower in the final state. The main backgrounds to the selection are: NC events, in particular those containing a $\pi^0$; inelastic $\nu_\mu$ CC events with little energy transferred to the muon; and the intrinsic beam $\nu_e$ contamination.

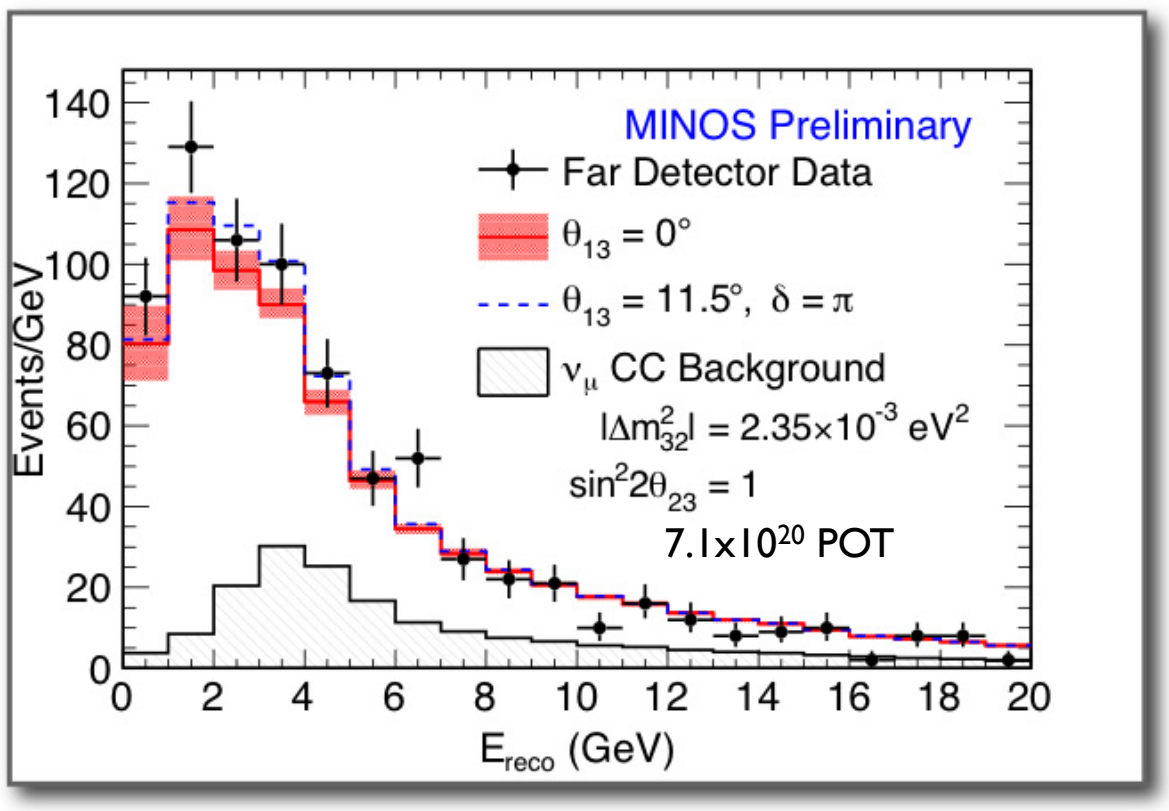

***Fig. 6:*** *Energy spectrum of the NC event candidates selected in the MINOS far detector. The data distribution (black points) is compared with predictions for standard three-flavor mixing, both with and without $\nu_e$ appearance (dashed and solid lines, respectively). The background from CC events is represented by the hatched histogram.*

The background prediction obtained from the near detector is $49.1 \pm 7.0(\text{stat.}) \pm 2.7(\text{syst.})$. MINOS selects 54 $\nu_e$ CC events in the far detctor, a 0.7σ excess over the expected background [18]. Figure 7 shows the possible values of the oscillation parameters that explain the data. By inverting the current of the NuMI horns and reversing the polarity of the detector magnets, MINOS operating conditions are optimized for antineutrino running. This enables MINOS to perform a comparison of the values of parameters governing oscillations of neutrinos with the values for antineutrino oscillations, expected to be identical in the standard picture. The methods employed for selection of the $\bar{\nu}_\mu$ CC sample are almost identical to the ones used in the previous MINOS CC analysis [10]. A total of 97 antineutrino events is observed in the far detector, with 155 antineutrino events predicted from the near detector observations. Hence, MINOS disfavors the null oscillation hypothesis at 6.3σ. The best fit values for the antineutrino oscillation parameters are $|\Delta\bar{m}^2_{32}| = 3.36^{+0.45}_{-0.40}(\text{stat.+syst.}) \times 10^{-3}\ \text{eV}^2$ and $\sin^2 2\bar{\theta}_{23} = 0.86 \pm 0.11(\text{stat.+syst.})$ [14]. The reconstructed antineutrino energy spectrum of the selected events is shown in Figure 8. The obtained 68% C.L. and 90% C.L. likelihood contours are displayed in Figure 9 and compared with the neutrino equivalents.

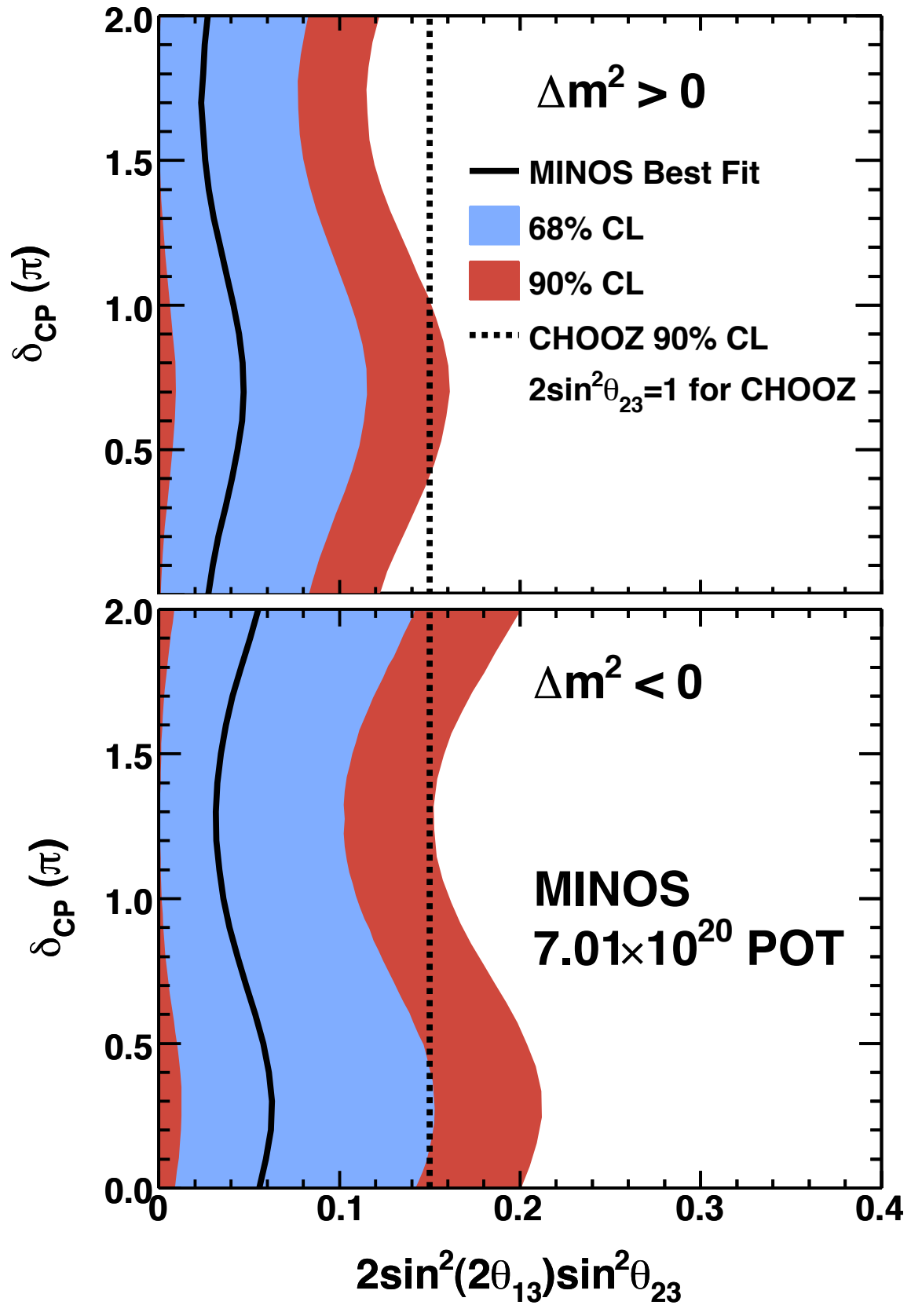

*Fig. 7: Values of oscillation parameters consistent with the observed data in the far detector for normal hierarchy (top) and inverted hierarchy (bottom). The black lines show the best fit values, whereas the red and blue regions show the 90% C.L. and 68% C.L. intervals, respectively. The CHOOZ limit, for* $\Delta m^2_{32} = 2.43\times10^{-3}$ eV$^2$, $\sin^2 2\theta_{23} = 1.0$, *is represented by the dotted line.*

A 2.3σ difference is observed between the best fit points. MINOS will proceed with further antineutrino running during 2011, with the aim of doubling the statistics of the antineutrino sample and clarifying the tension between the neutrino and antineutrino results.

## 3.4. The OPERA Experiment

The OPERA (Oscillation Project with Emulsion tRacking Apparatus) experiment aims to perform the first direct observation of $\nu_\mu \rightarrow \nu_\tau$ oscillations by looking for $\nu_\tau$ appearance in a $\nu_\mu$ beam. It uses the CERN to Gran Sasso (CNGS) beam, created from 400 GeV protons accelerated by the SPS facility, and measured with a far detector located 732 km away at the Laboratori Nazionali del Gran Sasso.

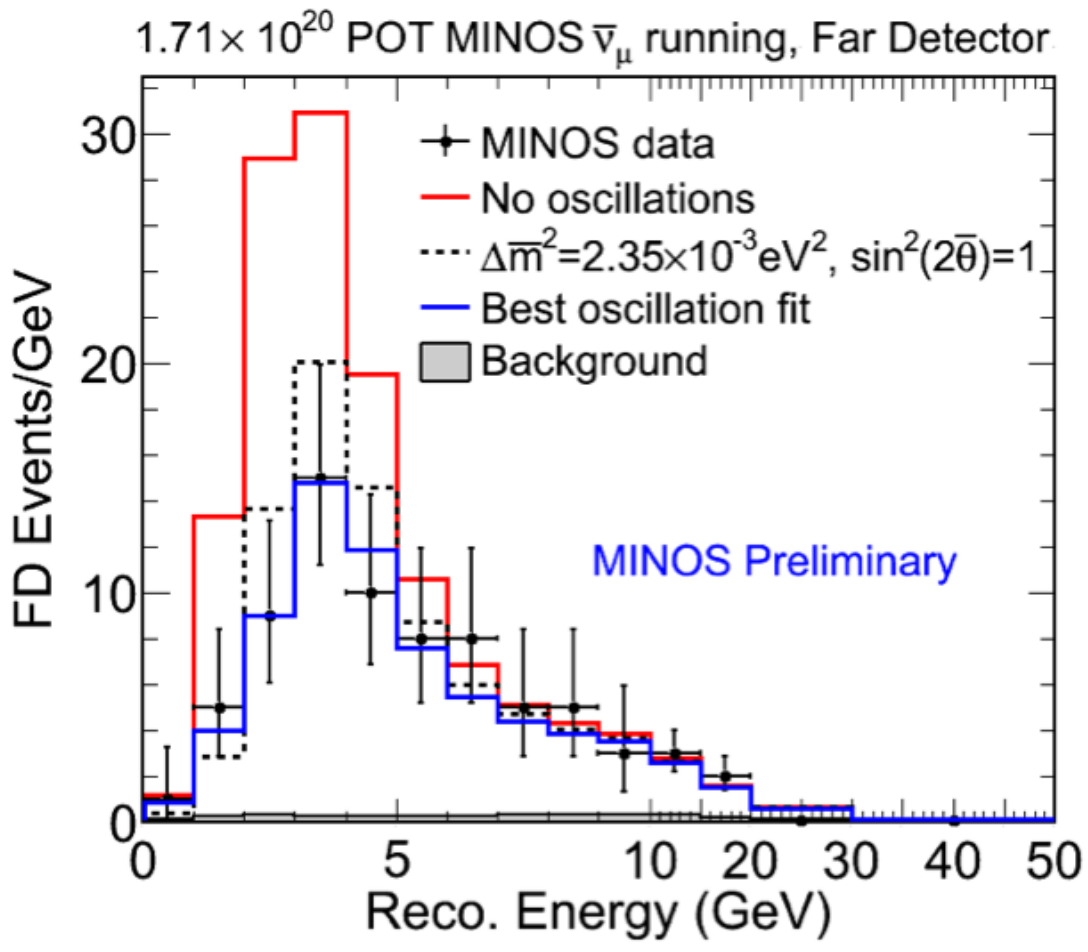

*Fig. 8: Reconstructed antineutrino energy of* $\bar{\nu}_\mu$ *CC selected events in the far detector. The black points are the data, the red histogram shows the prediction for no oscillations, the blue histogram shows the best fit to the data, and the dotted histogram represents the expectation assuming antineutrinos oscillate with the values obtained from the neutrino analysis.*

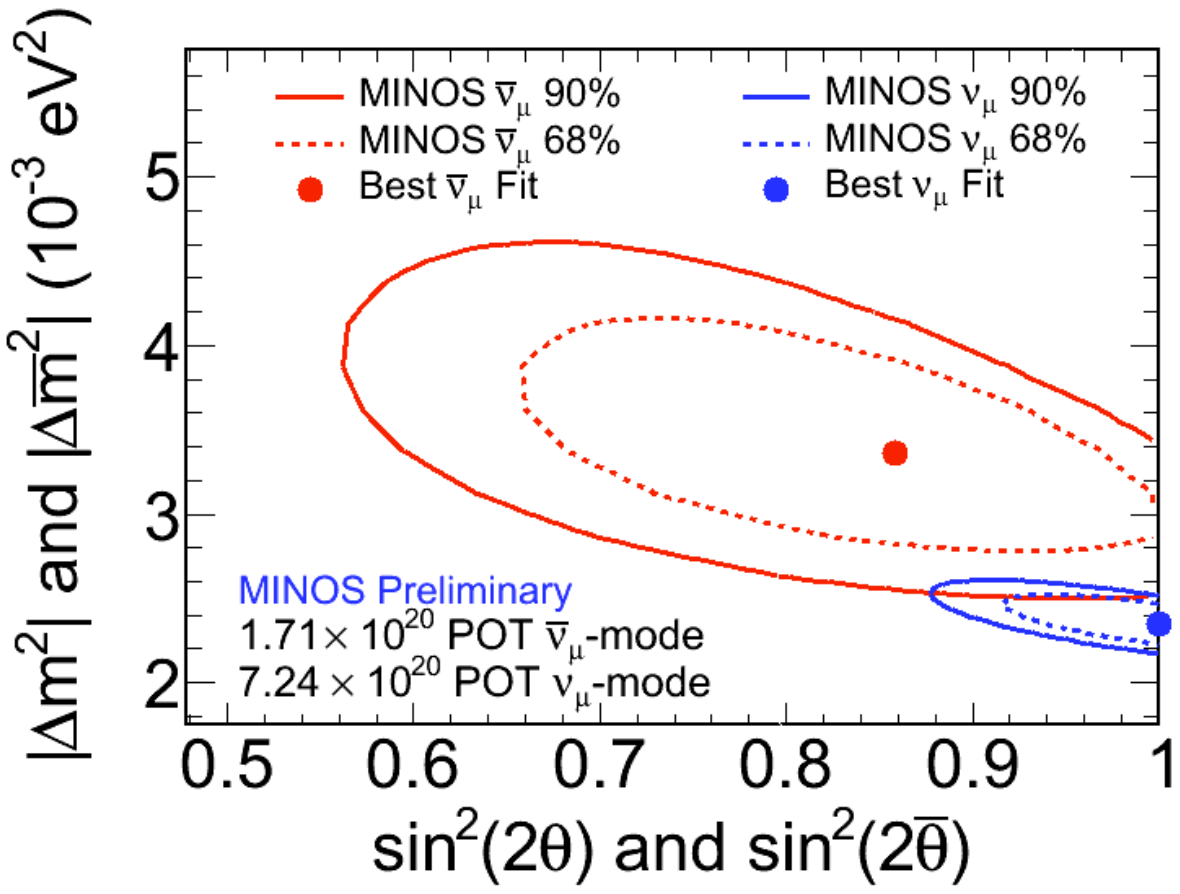

*Fig. 9: The MINOS 68% C.L. and 90% C.L. allowed regions for the antineutrino (red) and neutrino (blue) oscillation parameters.*

Most of the neutrinos in the beam have energies between 5 and 25 GeV energies, higher than the 3.5 GeV kinematic threshold for τ production, but also fairly distant from the oscillation maximum energy of 1.6 GeV for the 732 km baseline. Therefore, the expected number of $\nu_\tau$ interactions is small and the measurement is made more challenging by the difficulties of detecting neutrino-induced τ particles. This requires a complex detector composed of lead/emulsion bricks, which allow detection of the characteristic "kink" in tau decays, combined with scintillator trackers that help identify bricks containing candidate interactions. The bricks are robotically removed from the detector and the scanning of the emulsion sheets is fully automated. OPERA has collected over $7.0 \times 10^{19}$ POT in statistics since June 2008 and has already identified one $\nu_\tau$ event candidate. Further details on OPERA and its preliminary results are given elsewhere in these proceedings [19].

## 4. Next Generation Long-Baseline Experiments

The main goal of the next generation of long-baseline neutrino experiments, T2K and NOνA, is the observation of $\nu_e$ appearance and the measurement of the $\theta_{13}$ mixing angle. In order to reduce the large background to $\nu_e$ detection arising from NC interactions of high energy neutrinos, both experiments position their detectors offset from the neutrino beam axis. Due to pion decay kinematics, with a judicious choice of off-axis angle, the beam probed by the detectors can have a very narrow energy spectrum that peaks near the $P(\nu_\mu \rightarrow \nu_e)$ oscillation maximum. Figure 10 displays the NuMI beam spectra for different off-axis angles. The removal of high energy neutrinos from the beam spectrum drastically reduces the NC contamination in the $\nu_e$ CC selection.

### 4.1 The T2K Experiment

The Tokai to Kamioka experiment employs the new JPARC neutrino beam, created from 30 GeV protons, with a design intensity of 750 kW, to look for $\nu_e$ appearance over a 295 km baseline. The beam is aimed so that T2K's far detector, Super-Kamiokande's Čerenkov tank, is placed 2.5° off the beam axis, resulting in a narrow beam profile (0.3 GeV FWHM) peaked at the $P(\nu_\mu \rightarrow \nu_e)$ oscillation maximum of 0.6 GeV at 295 km. The T2K near detector complex contains two main detectors, INGRID, placed on-axis to measure beam profile and direction, and ND280, placed 2.5° off-axis and 280 m downstream of the target, to measure the beam flux, cross sections and beam composition.

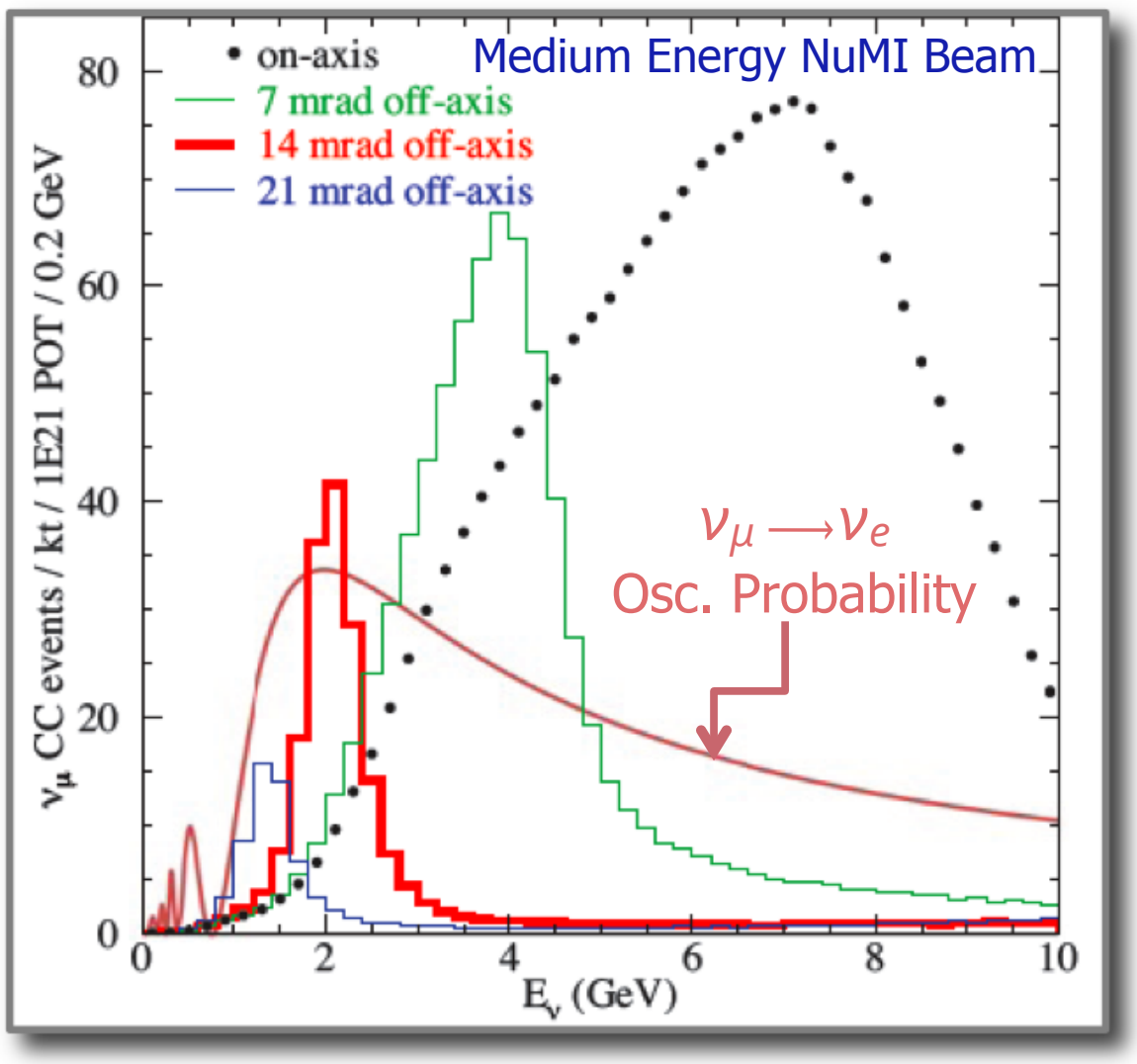

*Fig. 10: NuMI beam energy spectra for different off-axis angles. The red line corresponds to the configuration adopted by the NOνA experiment. For comparison, the $\nu_\mu \rightarrow \nu_e$ oscillation probability for an 810 km baseline is shown in arbitrary units as a function of neutrino energy.*

The ND280 detector is divided in three components, a tracker, a $\pi^0$ detector, and an electromagnetic calorimeter. The tracker is composed of three time projection chambers with two fine-grained scintillator detectors in between, and will measure the beam flux and energy spectrum, as well as perform neutrino cross section measurements, in order to constrain the far detector prediction. The $\pi^0$ detector consists of layers of triangular scintillator bars interleaved with water targets, and will study $\pi^0$ production in NC interactions, in order to understand the main background to the $\nu_e$ appearance measurement. T2K's run plan calls for 5 years of data taking with 750 kW beam intensity ($\sim 5 \times 10^{21}$ POT), to allow improvement of $\theta_{13}$ sensitivity by an order of magnitude over the CHOOZ limit. Figure 11 shows T2K's expected $\theta_{13}$ sensitivity. T2K has started operations and JPARC has delivered $3.4 \times 10^{19}$ POT. The beam has been successfully operated up to 100 kW beam power and ramping up to 750 kW will continue through 2011 and 2012. The first T2K events seen in the Super-Kamiokande detector were presented during the Summer 2010, showing that all components of the experiment are performing well.

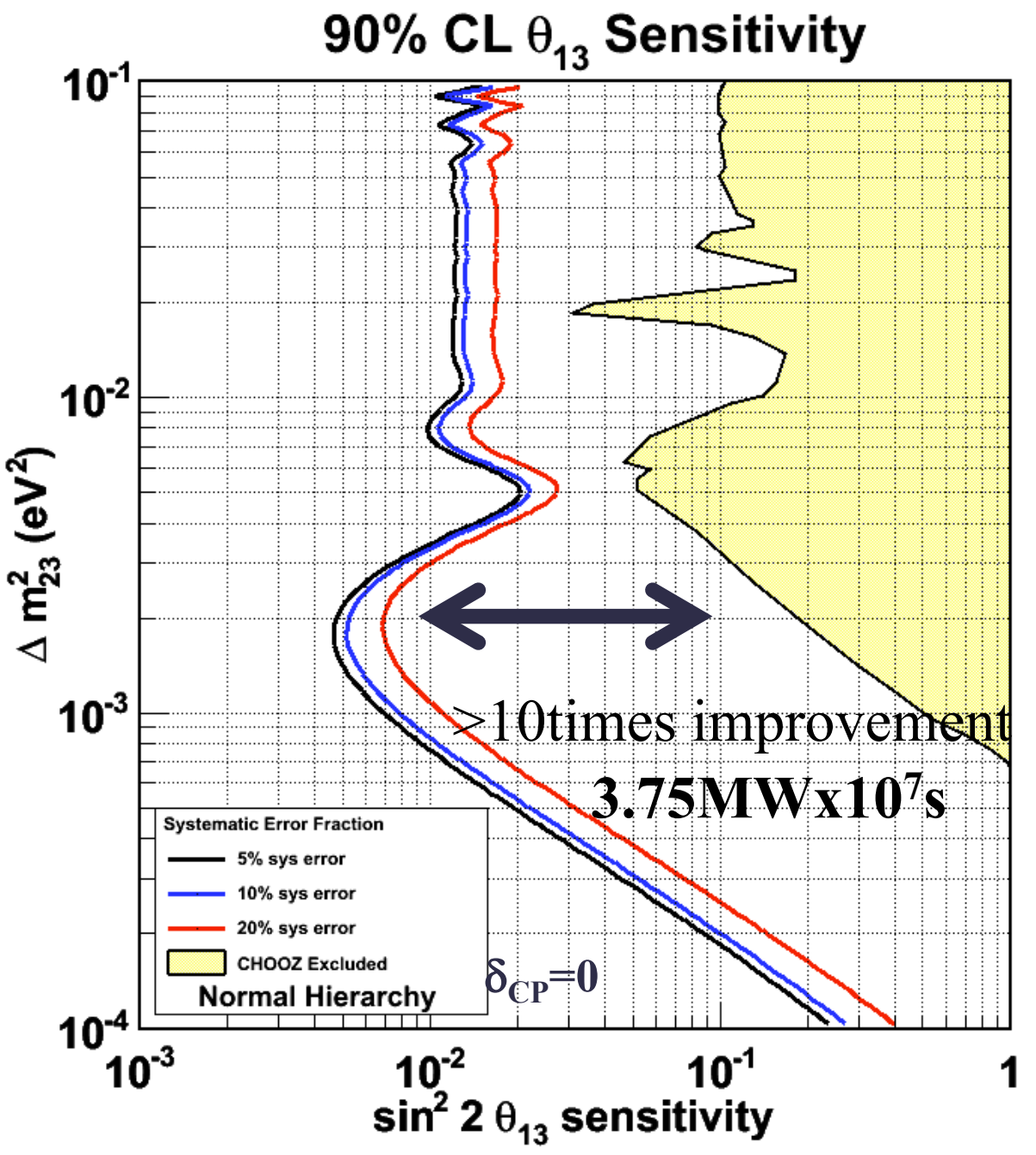

*Fig. 11: Expected T2K 90% C.L. sensitivity contours for the* $\sin^2 2\theta_{13}$ *measurement as a function of* $\Delta m^2_{32}$ *assuming 0.75 MWx5 year running. The yellow region shows the CHOOZ-excluded oscillation parameter values.*

## 4.2 The NOνA Experiment

Like T2K, the NuMI Off-axis $\nu_e$ Appearance experiment has the measurement of the $\theta_{13}$ mixing angle as its main goal. It will use the NuMI beam upgraded to 700 kW beam intensity and a large far detector, located 810 km away, 14 mrad (0.8°) off-axis, in Ash River, northern Minnesota. The narrow-band beam peaks at 2 GeV, near the $P(\nu_\mu \to \nu_e)$ oscillation maximum at 810 km. The 14 kton far detector and 0.22 kton near detector (placed 1km downstream of the target, also 14 mrad off-axis) share the same basic design. Both consist of planes composed of highly reflective PVC cells filled with liquid scintillator and read out by 32-pixel avalanche photodiodes. The detectors are optimized for detection of $\nu_e$ CC interactions with fine sampling of the electromagnetic showers (1 plane~0.15 $X_0$, Molière radius=10 cm). The NOνA physics program requires running for three years with a $\nu_\mu$ beam and three years with a $\bar{\nu}_\mu$ beam. Figure 12 shows the sensitivity to nonzero $\theta_{13}$ as a function of $\delta_{CP}$, which is one order of magnitude below the CHOOZ limit.

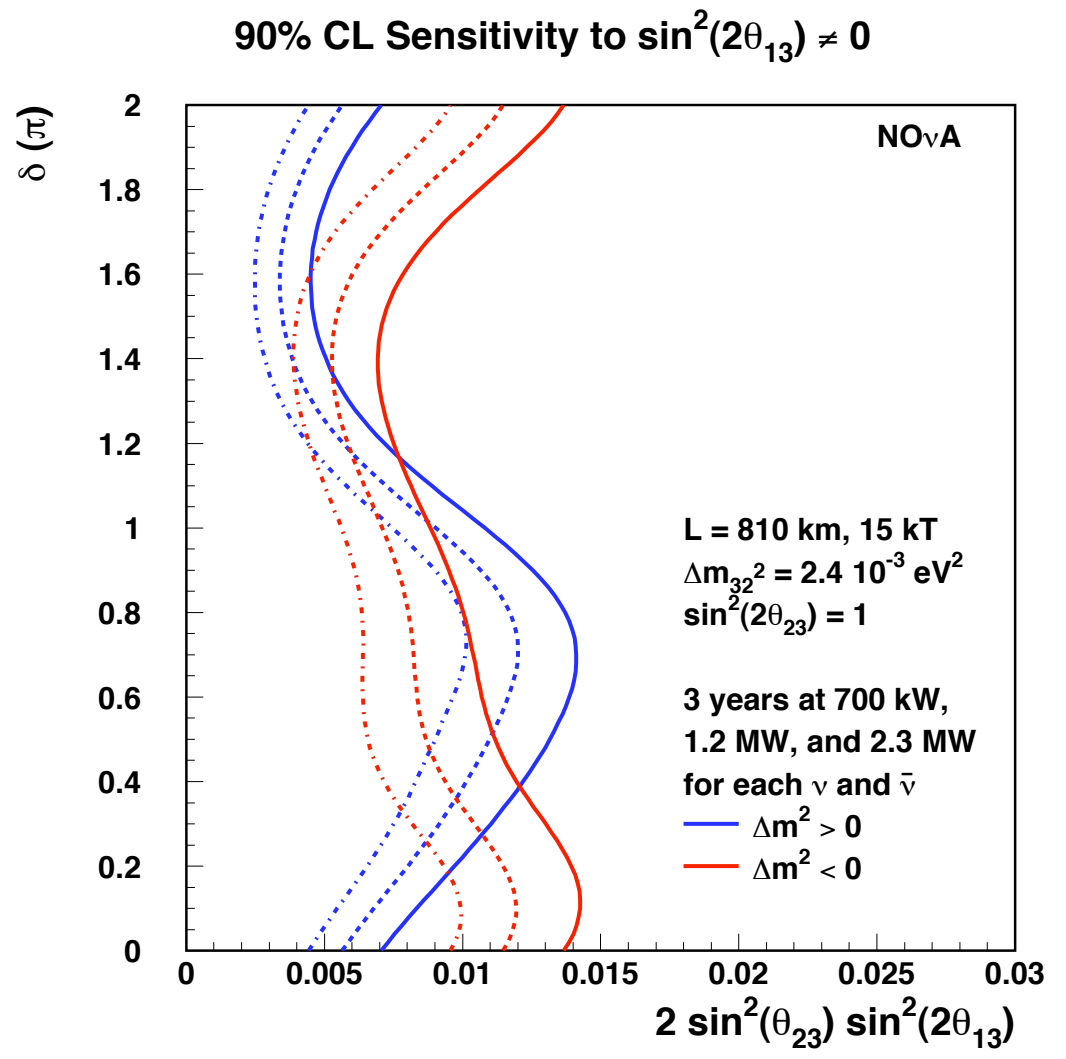

*Fig. 12: NOvA's sensitivity to* $\nu_e$ *appearance as a function of* $\delta_{CP}$ *and the mass hierarchy. The regions to the right of the curves are excluded at the 90% C.L.*

Due to NOνA's long baseline, matter-induced oscillations affect $\nu_\mu \to \nu_e$ and $\bar{\nu}_\mu \to \bar{\nu}_e$ oscillation probabilities by ~30%, granting NOνA the unique ability of resolving the mass hierarchy if $\theta_{13}$ is sufficiently large. Figure 13 shows NOνA's ability to resolve the mass hierarchy. NOνA has already begun data taking with the near detector placed on the surface. These data will be used for testing the detector components, as well as for calibration of the detector response and tuning of the Monte Carlo simulations using cosmic-ray and neutrino-induced events. Construction at the Ash River site is well underway, with far detector assembly starting in the Fall of 2011. Full operations using the upgraded NuMI beam, the completed far detector and the near detector underground are programmed to start in the Fall of 2013.

## 5. Future Experiments

The future generation of long-baseline experiments, currently in the R&D stages, is poised to improve even further the sensitivity to $\theta_{13}$, pin down the value of $\delta_{CP}$, and unambiguously determine the neutrino mass hierarchy. Different combinations of values of the mixing parameters can produce the same effects in the $\nu_\mu \rightarrow \nu_e$ oscillation probability, leading to ambiguities.

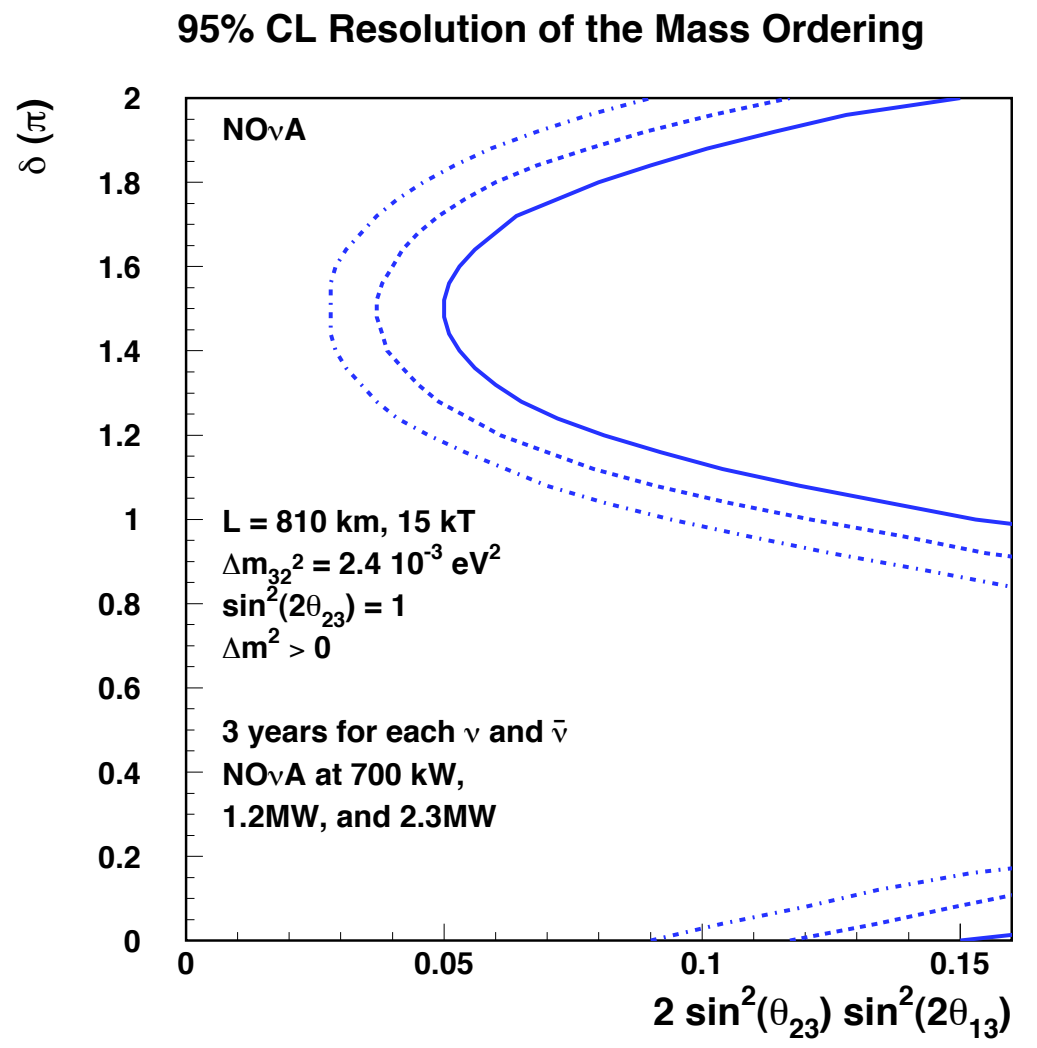

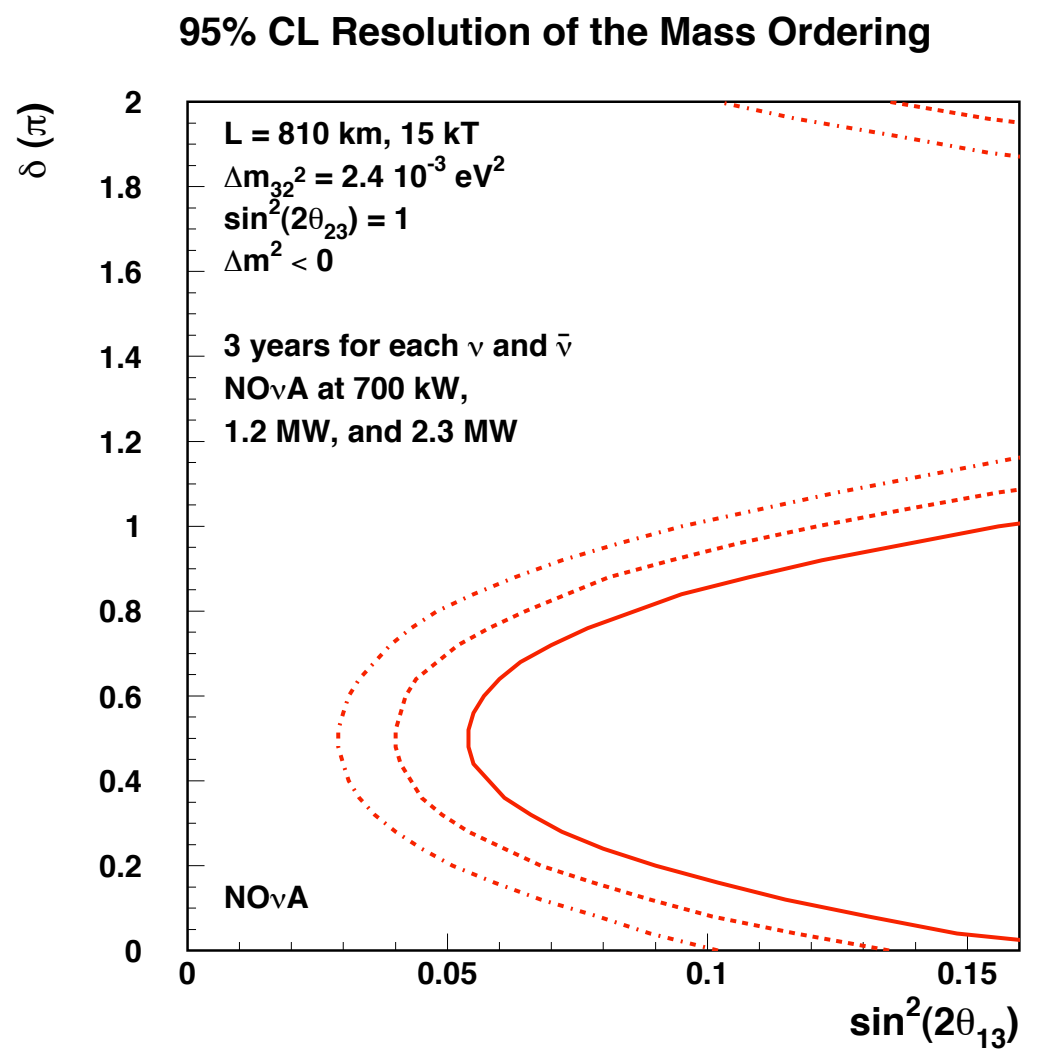

*Fig. 13: NOvA's ability to resolve the neutrino mass hierarchy if nature has chosen normal (top) or inverted (bottom) hierarchy. NOvA can resolve the mass hierarchy at 95% C.L. in the regions of parameter space to the right of the curves.*

The degeneracies can be resolved by measurements at different baselines and/or at different energy values. Achieving such precise measurements requires yet more intense beams and larger detectors. The LBNE (Long Baseline Neutrino Experiment) project, the flagship of the intensity frontier initiative at Femilab, is addressing these issues by designing a new neutrino beam line spanning 1300 km from Fermilab to DUSEL (Deep Underground Science and Engineering Laboratory) [20]. An initial beam power of 0.7 MW is envisaged, with upgrades to 1.2 MW and 2.3 MW possible. The 1300 km baseline affords sensitivity to the first two oscillation maxima, which for the normal mass hierarchy would be at neutrino energies of 2.4 and 0.8 GeV. This approach provides an avenue to lift the degeneracies mentioned above, but sensitivity to both maxima requires an on-axis wide-band beam resulting in increased backgrounds. LBNE is studying two detector technologies for $\nu_e$ CC detection: a fully active fine-grained liquid argon time projection chamber (LAr-TPC), with a mass of 50 kton; and a very large deep underground water Čerenkov detector with a mass of 300 kton. The possibility of a hybrid detector composed of a 20 kton LAr-TPC and a 100 kton water Čerenkov module is also being considered. Figure 14 shows the expected ability of LBNE to pin down the value of $\delta_{CP}$ as a function of $\sin^2 2\theta_{13}$. LBNE operations are expected to begin near the end of the present decade.

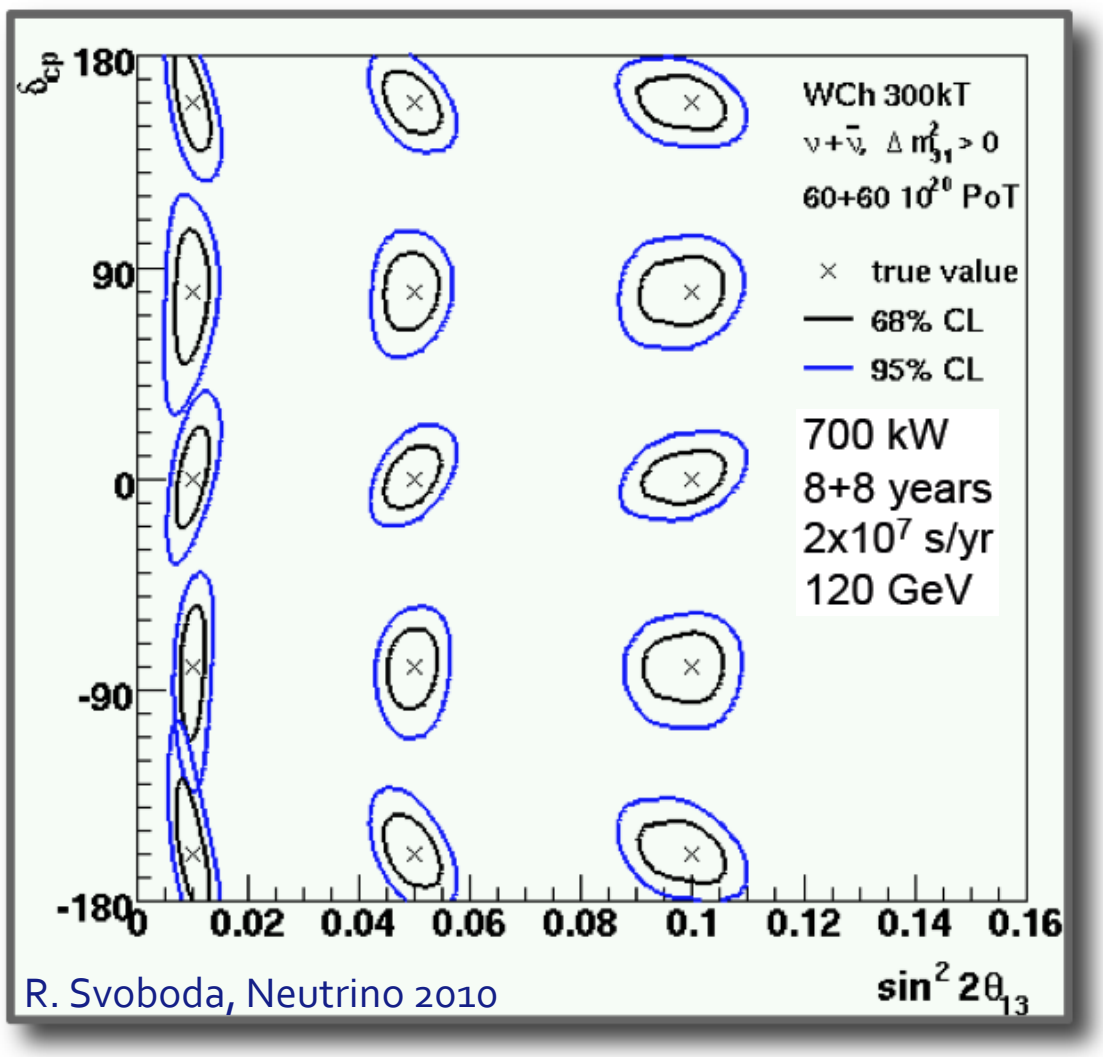

*Fig. 14: LBNE's ability to constrain $\delta_{CP}$ for different input values of $\delta_{CP}$ and $\sin^2 2\theta_{13}$. A neutrino beam with 700 kW beam power and a 300 kton water Čerenkov detector are assumed.*

In Japan, the T2KK (Tokay to Kamioka and Korea) proposes to measure the JPARC beam upgraded to 4 MW with two 270 kton water Čerenkov detectors, one located in Kamioka (HyperK) 295 km away, and the other located 1000 km away in South Korea. Using two different baselines with the same neutrino beam allows T2KK to lift the degeneracies between mixing parameters. A proposal for a 100 kton LAr-TPC detector in the Okinoshima island, with a baseline of 658 km, is also under consideration [21].

In Europe, the LAGUNA [22] and EUROnu [23] projects are studying different possibilities for a large facility to study neutrino oscillations. Water Čerenkov (MEMPHYS), LAr-TPC (GLACIER), and liquid scintillator (LENA) detector proposals are being studied. In parallel, several baselines radiating from CERN and ranging between 130 km to the Fréjus tunnel and 2300 km to the Pyhäsalmi mine are under review. The possible technologies for a new neutrino beam include a 4 MW super beam, a beta beam using heavy ions or a neutrino factory. Recommendations on the detector, site, and type of neutrino beam to be adopted are expected within the next 3 years.

## 6. Conclusions

Over the last decade, long-baseline neutrino oscillations experiments have greatly improved our knowledge of neutrino properties, solidly establishing neutrino mixing between active neutrino flavors, and ushered in the era of precision neutrino physics. The next generation experiments, focused on measuring $\theta_{13}$, may also begin constraining $\delta_{CP}$, and could resolve the neutrino mass hierarchy. Ongoing R&D and planning efforts for three new high-intensity neutrino facilities using very large detectors will address the challenge of fully resolving ambiguities among oscillation parameters.

Long-baseline experiments will continue to lead the way in unraveling the elusive nature of the neutrino, and assure us of very exciting times ahead in neutrino physics.